\documentclass[a4paper]{jpconf}
\usepackage{graphicx}

\newcommand{\be}{\begin{equation}} 
\newcommand{\en}{\end{equation}}
\newcommand{\bea}{\begin{eqnarray}}
\newcommand{\ena}{\end{eqnarray}}

\newcommand{\hbo}{\hbox to 1 true cm {\hfill } }

\begin{document}
\title{The density of states approach for the simulation of finite
  density quantum field theories } 

\author{\underline{K Langfeld$^a$}, B Lucini$^b$, A Rago$^a$, R
  Pellegrini$^c$ and L Bongiovanni$^b$}

\address{$^a$School of Computing \& Mathematics, Plymouth University,
  Plymouth PL4 8AA, UK}
\address{$^b$Department of Physics, Swansea University, Swansea SA2 8PP, UK}
\address{$^c$Higgs Centre for Theoretical Physics, University of
  Edinburgh, Edinburgh EH9 3JZ, UK} 

\ead{kurt.langfeld@plymouth.ac.uk}

\begin{abstract}
Finite density quantum field theories have evaded first principle Monte-Carlo
simulations due to the notorious sign-problem. The partition function of such
theories appears as the Fourier transform of the generalised density-of-states,
which is the probability distribution of the imaginary part of the action. With
the advent of Wang-Landau type simulation techniques and recent advances [1],
the density-of-states can be calculated over many hundreds of orders of
magnitude. Current research addresses the question whether the achieved
precision is high enough to reliably extract the finite density partition
function, which is exponentially suppressed with the volume. In my talk, I
review the state-of-play for the high precision calculations of the
density-of-states as well as the recent progress for obtaining reliable results
from highly oscillating integrals. I will review recent progress for
the $Z_3$ quantum field theory for which results can be obtained from the
simulation of the dual theory, which appears to free of a sign problem. 

\end{abstract}

\section{Introduction}

\begin{figure}[h]
\includegraphics[width=0.8\textwidth]{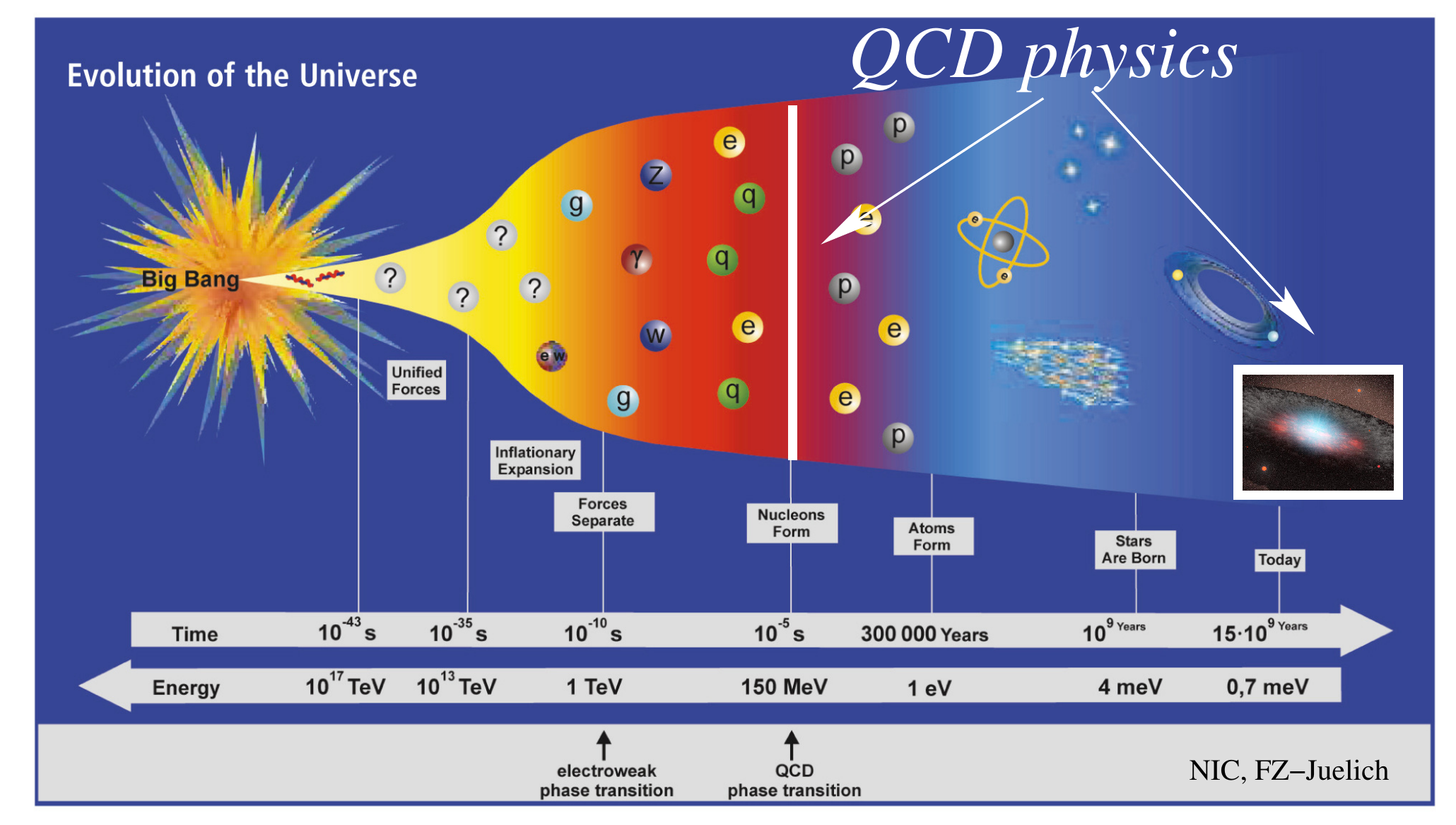}
\caption{\label{fig:1} QCD impact on the evolution of the early
  Universe. }
\end{figure}

\begin{figure}[h]
\begin{minipage}{15pc}
\includegraphics[width=15pc]{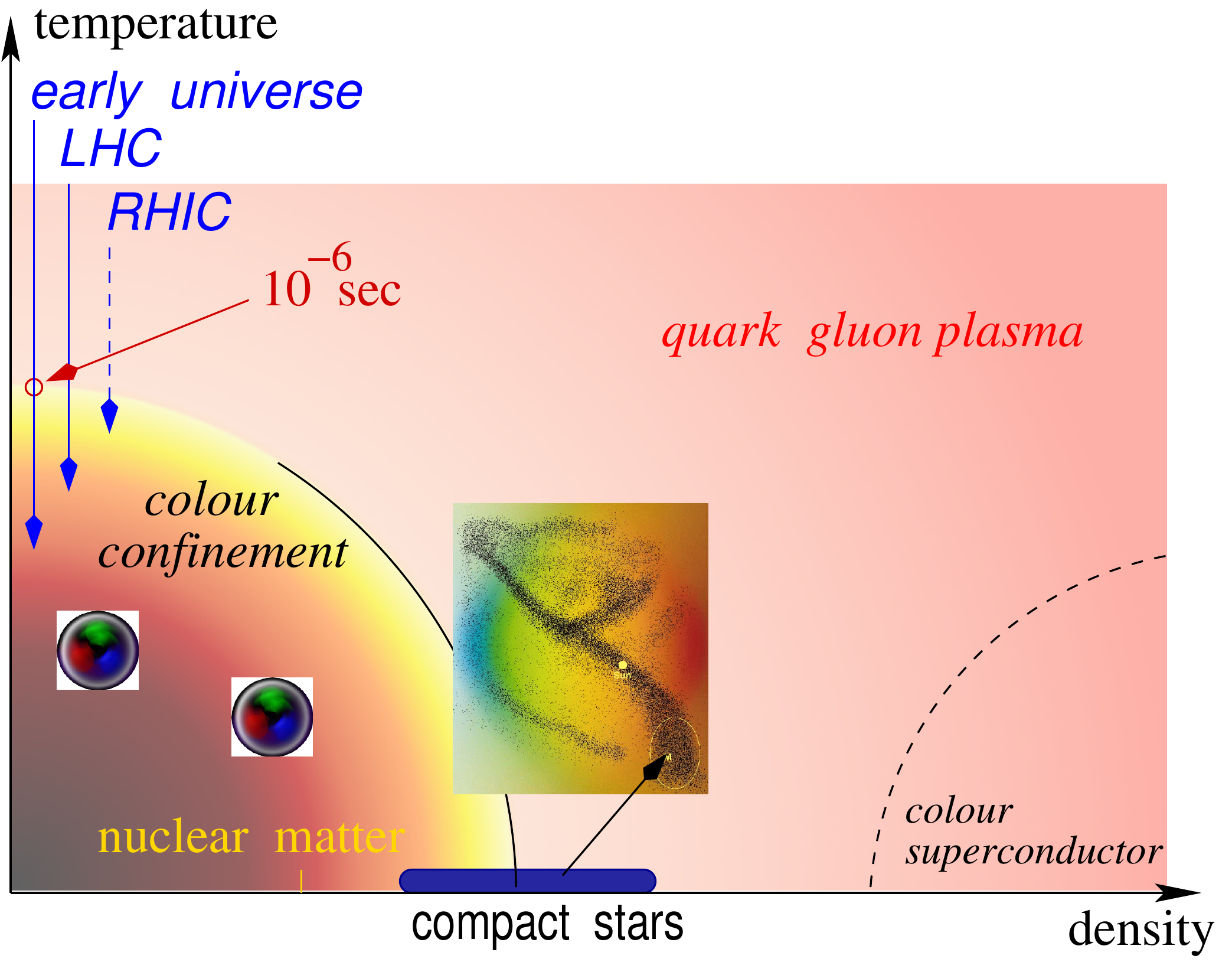}
\caption{\label{fig:2a} Sketch of the QCD phase diagram. }
\end{minipage}\hspace{2pc}%
\begin{minipage}{15pc}
\includegraphics[width=15pc]{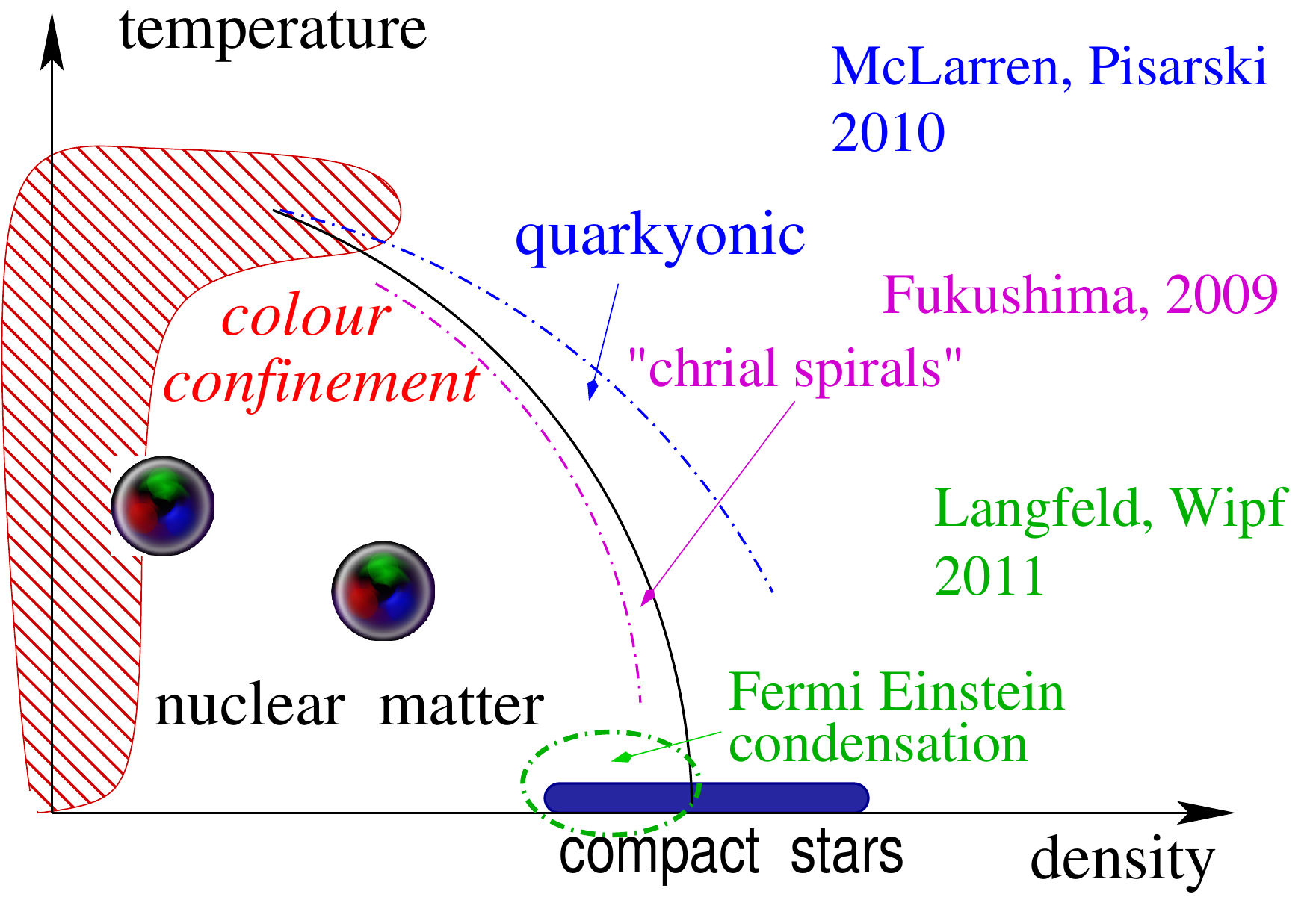}
\caption{\label{fig:2b} QCD phase diagram: first principle
  calculations and model building. }
\end{minipage} 
\end{figure}
The properties of strongly interacting matter has sparked many
important investigations using accelerator experiments and large scale
theoretical studies. In an astrophysical context, the theory of
interactions, QCD, dominates the area around  $10^{-6}$ seconds after
the Big Bang when quark matter confines to colour neutral
hadrons. Since QCD captures the impact of the strong nuclear forces,
it is widely believed that QCD plays an essential role to understand
compact star matter as it may exist nowadays e.g.~in neutron stars
(see figure~\ref{fig:1} for an illustration). The so-called QCD phase 
diagram characterises the states of matter as a function of the
density and temperature in a 2d graph. Completing this diagram in its
extreme regions and in particular for cold and dense matter is an
outstanding problem which still triggers model building and new
techniques for computer simulation after 30 years of intense
research. Under moderate conditions, quarks and gluons are confined to
colour-neutral hadrons, while this {\it confinement} feature of QCD is
believed to cease to exist under extreme conditions, temperatures
and/or densities. In the early universe before the QCD phase
transition roughly at time $10^{-6}$ seconds, matter has not yet
clustered due to gravity and, as a result, the density is very low
compared to e.g.~nuclear matter density. Similarly, the conditions in
heavy ion collision experiments carried out at RHIC (located at
Brookhaven National Laboratory (BNL) in Upton, New York) or LHC (built 
by the European Organization for Nuclear Research (CERN)) produce very
hot matter at low densities. From an experimental point of view, very
little is known even at moderate densities let alone the cold and
dense regime of compact star matters (see figure~\ref{fig:2a}). 

\medskip
Lattice gauge simulations offer first principle non-perturbative
results with good control over the errors. Markov chain Monte-Carlo
simulations can be very successfully applied and reliable results for
the states of matter can be obtained at all temperatures and small
baryon chemical potentials  (shaded area in  figure~\ref{fig:2b}, see
e.g.~\cite{Laermann:2003cv} for a review). Over the recent years, many
new models have been developed to describe QCD matter at high
densities: quarkyonic matter is motivated by the so-called 't~Hooft
limit of the hypothetical theory with a large number of colours and
sees the quarks deconfined inside the Fermi sphere while only the
baryons at the surface of the Fermi sphere undergo colour 
confinement|\cite{McLerran:2007qj}. The chiral magnetic
effect~\cite{Fukushima:2010fe} takes into account the strong magnetic
field that are produced by the colliding charges in heavy ion
experiments and uses an effective quark model to estimate their impact
on the QCD phase structure. In the confinement phase, quarks in a
certain gluonic background can change their statistics from Fermi to
Bose type. At moderate densities, these so-called centre dressed and
confined quarks therefore might undergo Bose condensation leading to to
new state of cold and confined matter~\cite{Langfeld:2011rh}. None of
these ideas have yet been tested by first principle
calculations. At a finite baryon chemical potential, the QCD action
acquires an imaginary part, and, hence, standard Monte-Carlo
techniques cannot be applied for the lattice simulation of QCD at
finite densities. This has become known as the {\it notorious sign
  problem.} Early attempts pursued Monte-Carlo simulations with
the modulus of the quark determinant and included the phase factor to
the observable to be calculated. It turns out that the expectation
value of the phase factor is very small (see~\cite{Splittorff:2006fu}
for an illustration)  showing that the such generated Monte-Carlo
configurations contain very little information on finite density
QCD. The {\it sign} problem has turned into an {\it overlap} problem. 

\medskip 
The recent past has seen a renaissance of ideas targeting dense and 
cold fermionic matter.  Some of the methods were proposed some time
ago, but have now reached an unprecedented level of sophistication. The
{\it reweighting} approach proposed by Fodor and
Katz~\cite{Fodor:2001au} uses a multi-parameter action to optimise the
overlap. This method is particularly suited for intermediate
temperatures and might possess a reach that covers the critical
endpoint of the QCD phase diagram~\cite{Fodor:2001pe}. Langevin
simulations of lattice gauge theories avoid the positivity constraint
of the Gibbs factor, which lies at the heart of Monte-Carlo
simulations, and might therefore be suitable for finite density
simulations~\cite{Parisi:1984cs,Karsch:1985cb}. This technique
regained a lot of interest when Aarts showed that {\it stochastic
  quantisation} can evade the sign problem at least for the
relativistic Bose gas~\cite{Aarts:2008wh,Aarts:2009hn}. Although the
conceptional question whether the approach converges to the correct
answer~\cite{Aarts:2009uq} is still under active investigations, the
approach is one of the few methods that are currently applied to
finite density QCD~\cite{Aarts:2014bwa}. It might appear that
integrating the gluonic degrees of freedom {\it before} the fermion
fields  alleviates the sign problem. This could be done e.g.~in the
strong coupling limit~\cite{Rossi:1984cv} leading to a description of
Nuclear Physics suiting lattice simulations~\cite{deForcrand:2009dh}. 
It was observed in 1d QCD that even integrating part of the gluonic
degrees of freedom leads to substantial
improvements~\cite{Bloch:2013ara}. Finally, a reformulation of the
theory might improve on the sign the problem or remove it
altogether. Indeed, theories the dual of which are real are then
accessible by standard Monte-Carlo techniques. Example for 
complex action spin models that are real upon dualisation are $Z_3$
spin model~\cite{Mercado:2011ua} or the $O(2)$ model at 
finite densities~\cite{Banerjee:2010kc,Langfeld:2013kno}. These models
can be efficiently simulated using worm type
algorithms~\cite{Prokof'ev:2009xw}.  Alternative lattice
discretisations~\cite{Brandt:2014rca} and spin blocking techniques in
combination with the (tensor) renormalisation group
approach~\cite{Meurice:2014tca} might be equally successful to
eliminate the sign problem from Yang-Mills theories. Also not hinging
on a real dualisation  of the theory is the {\it fermion bag approach}
approach by Chandrasekharan~\cite{Chandrasekharan:2009wc} for which
the sign problem is relegated to finite size fermion bags. This
approach has been seen to be very efficient for fermion theories with
four-fermion  coupling such as the Thirring model with massless
fermions on large lattices~\cite{Chandrasekharan:2011mn}.  

\medskip 
An efficient alternative to conventional Monte-Carlo simulations is
based upon the numerical computation of the density of states using
the multi-canonical algorithm~\cite{Bazavov:2012ex} or a Wang-Landau
type strategy~\cite{WangLandau2001}. A modified version of the
Wang-Landau method is the LLR algorithm~\cite{Langfeld:2012ah}, which
is effective for theories with continuous degrees of freedom as
opposed to spin models (see also~\cite{Pellegrini:2014gha}). The
latter algorithm has been extended from the calculation of the action
distribution to accessing probability distributions of other extensive
quantities such as the SU(2) Polyakov
line~\cite{Langfeld:2013xbf}. Furthermore, it has been proposed
in~\cite{Langfeld:2014nta} to use LLR techniques for a high precision 
calculation of the distribution of the {\it imaginary part} of the
action. Once this quantity has been determined, the partition function 
of the complex theory can be computed semi-analytically by carrying
out the Fourier transform of the corresponding probability
distribution~\cite{Langfeld:2014nta}. Below, we will summarise the
state of affairs concerning density-of-states methods and the LLR
algorithm in particular to simulate theories with a sign problem. An
overview on selected new methods solving the sign problem can be 
found in the recent review by Aarts~\cite{Aarts:2015kea}.

\section{Density-of-states approach to complex action systems}

\subsection{Density-of-states and the overlap problem}

Let us consider the partition function $Z$ of a theory of one degree
of freedom $\phi $ with a complex action: 
\be 
Z = \int {\cal D}\phi \; 
\exp \{ \beta S_R[\phi] +  i \mu S_I[\phi]  \} 
\label{eq:1}
\en 
where $\mu $ is the ``chemical potential'', and $S_{R/I} $ are real
and imaginary parts of the action. We introduce the generalised density
of states~\cite{Langfeld:2014nta} by 
\be 
P_\beta (s) = \int {\cal D}\phi \; \delta \Bigl( s - S_I[\phi] \Bigr) \; 
\exp \{ \beta S_R[\phi]\} \; . 
\label{eq:2}
\en 
For $\beta=0$, $P_0(s)$ just counts the number of states with the
constraint that the imaginary part of the action is given by $s$. At
finite $\beta $, the number count is weighted by the ``Gibbs'' factor
$ \exp \{ \beta S_R[\phi]\}$. Once $P_\beta (s)$ is known, the task to
calculate the partition function boils down to evaluate the integral
\be 
Z(\beta,\mu) \; = \; \int ds \; P_\beta (s) \; \exp \{ i \,  \mu  \,
s\} . 
\label{eq:3}
\en
Since the integrand in (\ref{eq:2}) is perfectly real, the
difficulties with the sign problem are relegated to the 1-dimensional
integral (\ref{eq:3}). In fact, $P_\beta (s)$ could be estimated by
performing a standard Monte-Carlo simulation with the Gibbs factor 
$\exp \{ \beta S_R[\phi]\}$ and to bin the values for $S_I$ in a
histogram. The LLR algorithm will provide us with $\beta=0$, $P_0(s)$
over hundreds of orders of magnitude (see below for an example). The
aim of this paper is to discuss the options for a calculation of the
highly oscillating integral (\ref{eq:3}). 

\medskip 
Let us scope the amount of difficulty that resides with this task. We
firstly note that the action $S_i$ is an extensive quantity $S_I(\phi)
= V s_I$ with $V$ the number of degrees of freedom (volume) and with the
action density $s_i$ of order one. A good qualitative choice (see
e.g.~\cite{Langfeld:2013kno}) is given by 
$$
P_\beta (s) = \exp \Bigl\{ - \frac{ s^2 }{ V } \Bigr\} 
\hbo \hbox{leading to } \hbo 
Z \; = \; \int ds \; \mathrm{e}^{-s^2/{ V}} \; \exp \{ i \, \mu s \} 
\; \propto \; { \exp\{ - \frac{\mu^2}{4} \, { V} \} } \; . 
$$
For a chemical potential $\mu $ of order one, $Z$ is exponentially
suppressed with the number of degrees of freedom $V$. On the other
hand, $P_\beta (s)$ is only known numerically and of order one at
least for small $s$. For a successful evaluation of the integral
in $Z$ (\ref{eq:3}), any numerical method for obtaining  $P_\beta (s)$
must have the properties
\begin{itemize}
\item exponential error suppression for extensive quantities
\item for the whole domain of support for $S_I$. 
\end{itemize}
The LLR algorithm proposed in~\cite{Langfeld:2012ah} just delivers
that. 

\subsection{The $Z_3$ spin model as showcase }

\begin{figure}[t]
\begin{minipage}{18pc}
\includegraphics[width=18pc]{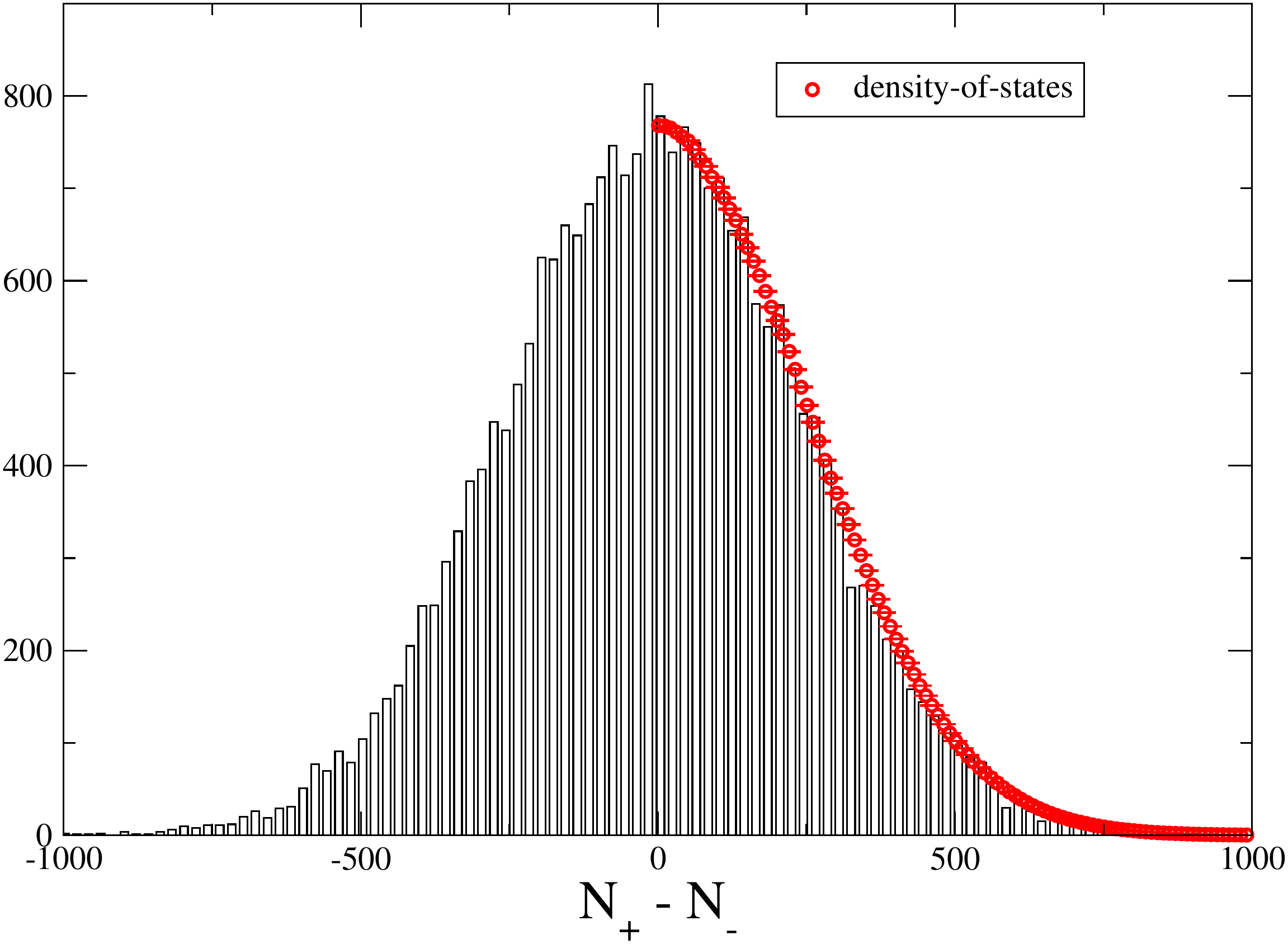}
\caption{\label{fig:3a} Histogram count for $N_+-N_- \propto S_I$
  (linear scale); $24^3$ lattice, $\tau=0.17$, $\kappa=0.05$. }
\end{minipage}\hspace{2pc}%
\begin{minipage}{14pc}
\includegraphics[width=14pc]{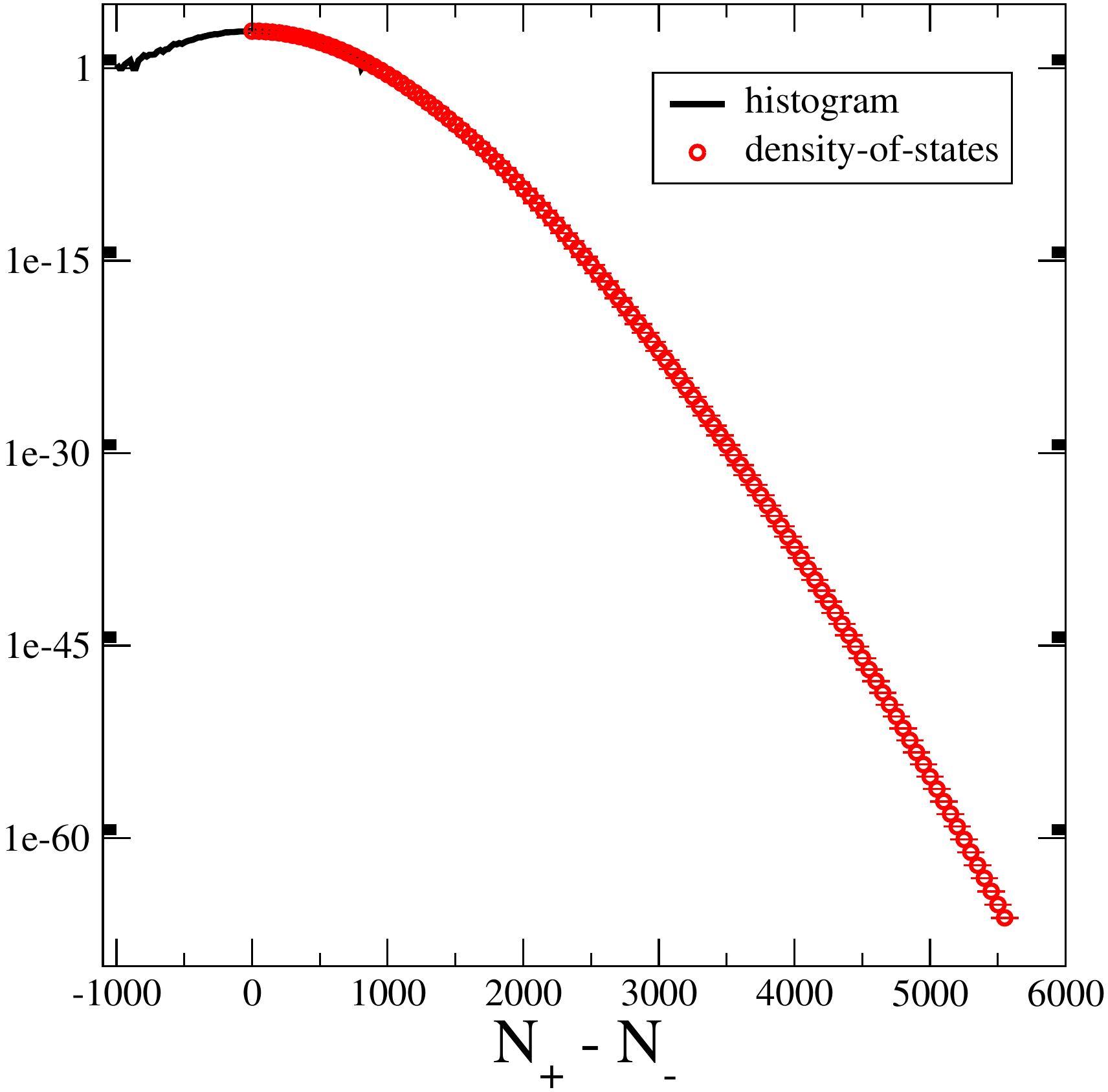}
\caption{\label{fig:3b} Same histogram on a logarithmic scale with the
  LLR result now reaching beyond $ N_+-N_- \approx 5,500$.  }
\end{minipage} 
\end{figure}
The key question is whether the quality of the result for $P_\beta
(s)$ obtained by the LLR algorithm is good enough to admit a reliable
calculation of the partition function $Z$ via the integral
(\ref{eq:3}). The answer to this question is model dependent. The 
3-dimensional $Z_3$ spin model on a cubic lattice at finite density
maps onto a real action system upon dualisation and is thus open to
standard Monte-Carlo simulations. It serves as a first benchmark test
in the feasibility study for our approach to theories with a sign
problem. Here, we will review our findings for the 3-dimensional
case. Degrees of freedom are the centre elements $z(x)$ that take
values from the group $Z_3$ 
$$ 
z(x) \; \in \{1,z_+,z_-\} \; , \hbo z_{\pm } = (1 + i \sqrt{3})/2 . 
$$
The action is given by 
\be 
S[z] = { \tau} \sum _{x,\nu} [z_x z^\ast _{x+\nu} +cc ] \; + \; 
\sum _x [ { \eta} z_x + { \bar{\eta}} z^\ast _x ] \; , \hbo 
\eta = \kappa \; \exp (\mu ) \; , \; \; \; \bar{\eta } = \kappa \;
\exp ( - \mu ) \; . 
\label{eq:10}
\en 
This model is inspired by finite density QCD in the heavy quark limit,
and the parameter $\tau $ is reminiscent of the temperature and
$\kappa $ reflects the quark hopping parameter~\cite{Mercado:2011ua}. 
Apparently, the action becomes complex as soon as $\mu \not=0 $. If
for a given configuration $z(x)$ the quantity $N_\pm$ represents the
number of spins on the lattice with $z= z_\pm$, the imaginary part of
the action can be written as: 
\be
S_I \; = \; \frac{\eta - \bar{\eta} }{2i} \sum _x [z \, - \, z^\ast]
\; = \; \sqrt{3} \, \kappa \, \mathrm{sinh}(\mu) \; [N_+ - N_-] \; . 
\label{eq:11}
\en
We have performed a standard Monte-Carlo simulation using a $24^3$
lattice, $\kappa = 0.17$ and $\kappa = 0.05$ to obtain a histogram for 
$N_+ - N_-$ (see~\cite{Langfeld:2014nta} for details).  The result is
shown in figure~\ref{fig:3a} on a linear scale (see
figure~\ref{fig:3b} with the $y$-axis on a logarithmic scale). Note
that we have very little ``events'' with $N_+- N_- >1000$. For the
calculation of the Fourier transform to obtain the partition function
$Z$ in (\ref{eq:3}), the tails with $\vert N_+-N_- \vert \gg 1000$
significantly contribute for $\mu \approx 1$. We recover the {\it
  overlap problem} in the light of the density-of-states
approach. Our result for $P(N_+-N_-)$ using the LLR method is also
shown in the figures~\ref{fig:3a} and \ref{fig:3b}. We find a
reassuring agreement with the standard simulation result and moreover
we have obtained the distribution $P(N_+-N_-)$ for values as large as 
$N_+- N_- =5000$ and over sixty orders of magnitude. 

\section{The partition function from highly oscillating
  integrals\label{sec:osc}}  

\subsection{Polynomial fit \label{sec:pf} } 

Our task is now to carry out the Fourier transform of the generalised
density of states $P (s)$ in order to obtain the partition
function $Z$ (see (\ref{eq:3})). One advantage of our approach is that
we can use sophisticated integration techniques, which converge like
$1/n^p$, $p>1$, where $n$ is the number 
of evaluations of the integrand. Note that any Monte-Carlo
integration that sub-sums the Fourier transform necessarily converges at best
like $1/\sqrt{n}$. Note, however, that even the sophisticated
integrations techniques would fail for sizeable values of the chemical
potential without further knowledge of the function $P (s)$. We
have so far studied the density-of-states for the theories $U(1)$,
$SU(2)$, $SU(3)$ and $Z_3$ and a common feature has been that 
$log \, P(s)$ is indeed  very smooth and, as expected, monotonic
functions of its variable $s$. In the present case, we also have the
symmetry under reflection $P (s) = P (-s)$. The ``smoothness'' of 
$P (s)$ is summarised by the fact that the Taylor expansion 
\be 
\ln \, P (s) \; = \; \sum _{i>0, \mathrm{even} }^q c_i \, s^i \;
, \hbo q =  2,4,6,8,\ldots 
\label{eq:20}
\en 
can produce results that are indistinguishable from the numerical
findings for $P (s)$ within error bars. Depending on the region
of the parameter space $\beta = (\tau,\kappa)$, $q$ as small as $4$ might be
sufficient. As soon as an acceptable representation of the numerical
data in terms of the ansatz (\ref{eq:20}) is found, the calculation of
the partition function can be performed in a ``semi-analytic'' way
using advanced numerical integration techniques: 
\be 
Z(\mu) \; = \; 2 \int ds \; P (s) \; \cos ( \mu \, 
\, s \, ) \; = \;  2 \int ds \; \exp \left\{ \sum _i^q c_i s^i
\right\} \,  \cos (\mu \, s ) \; .  
\label{eq:21}
\en 

\subsection{Asymptotic referencing} 

Similar to the scenario in the previous subsection, it might be useful
to describe the gross features of $P(s)$ by an analytic function, say
$P_\mathrm{asy}(s)$ especially for large values $s$. Decomposing 
\be 
P(s) \; = \; \bar{P}(s) \, P_\mathrm{asy}(s) \; , 
\label{eq:22}
\en 
the function $ \bar{P}(s) $ might have a moderate range of values
although $P(s)$ spans many orders of magnitude. For a technical
side-remark, we point out that the LLR
algorithm~\cite{Langfeld:2012ah} can be easily adapted to directly
produce $ \bar{P}(s) $ for a given choice for $P_\mathrm{asy}(s)$. 
The partition function is obtained by Fourier transformation: 
\be
Z(\mu) \; = \; \hbox{FT}[P](\mu) \; = \; \int ds \; P(s) \;
\mathrm{e}^{i\mu s} \; = \; \int dx \;  \hbox{FT}[\bar{P}](x) \;
\hbox{FT}[P_\mathrm{asy}](\mu \, - \, x )  \; . 
\label{eq:23}
\en 
The idea is that $\hbox{FT}[P_\mathrm{asy}]$ is analytically available 
and has already incorporated a good deal of the cancellations. For
moderate values of $\kappa $ and $\tau $, a sensible choice is 
\be
P_\mathrm{asy}(s) \; \propto \; \exp\{- \alpha s^2\} \hbo \hbox{leading to}
\hbo \hbox{FT}[P_\mathrm{asy}](\mu \, - \, x ) \; \propto \; 
\exp \left\{ - \, \frac{ (\mu-x)^2}{4 \alpha } \right\} \; . 
\label{eq:24}
\en 
Depending on the size of the intrinsic scale $\alpha $, we only need
to numerically calculate $\bar{P}(s)$ for $s \approx \mu $ for the 
chemical potential $\mu $ of interest.

\subsection{Eigenfunction expansion} 

Let us expand the generalised density-of-states $P(s)$ in terms of a
complete set of eigenfunctions (in the L2 sense) that are
eigenfunctions of a differential operator:  
\be 
P(s) \; = \; \sum _n c_n \; \psi _n(ks) \; , 
\label{eq:30}
\en 
where $k$ is a parameter that will be adapted to the intrinsic scale of
the theory under studies. We need not necessarily adopt an eigensystem
with an all discrete set of eigenfunctions. Here, we have indeed the
eigenfunctions of the harmonic oscillator in mind having this
property: 
\be 
- \; \frac{d^2}{ds^2} \psi _n (ks) \; + \; k^4 \, s^2 \; \psi _n(ks) \; = \; 
k^2 \; (2n+1) \; \psi _n (ks) \; .  
\label{eq:31}
\en
The ortho-normal eigenfunctions are the well-known Hermite functions: 
\be 
\psi_n(x) \; = \; \frac{1}{\sqrt{2^n n! \sqrt{\pi} }} \;
\mathrm{e}^{-x^2/2} \;  H_n(x) \; , \hbo 
H_n(x) \; = \; (-1)^n \; \mathrm{e}^{x^2} \; \frac{d^n}{dx^n} \; 
\mathrm{e}^{-x^2}  \; , 
\label{eq:32}
\en 
where the $H_n(x)$ are the Hermite polynomials. Using the
``Schr\"odinger equation'' (\ref{eq:31}), one proves that the
eigenfunctions are fixed points of the Fourier transformation: 
\be 
\hbox{FT}[\psi _n](\mu) \; = \; \int ds \; \psi_n (ks) \;
\mathrm{e}^{i\mu s} \; = \; \frac{ \sqrt{2 \pi}}{k} \; i^n \; 
\psi _n \left( \frac{\mu}{k} \right) \; . 
\label{eq:33}
\en 
We therefore find for the partition function 
\be 
Z(\mu) \; = \; \hbox{FT}[P](\mu) \; = \; \sum _n c_n \; \hbox{FT}[\psi
_n](\mu) \; = \;  \frac{ \sqrt{2 \pi}}{k} \; \sum _n i^n \; c_n \; \psi _n
\left( \frac{\mu}{k} \right) \; . 
\label{eq:34}
\en 
If the chemical potential $\mu $ is larger than the intrinsic scale
$k$, we observe that we attain small values for $Z$ already from the
asymptotic behaviour of the Hermite functions, i.e., $\psi _n(x) 
\approx \exp \{-x^2/2\}$. 

\section{The $Z_3$ spin model - numerical results}

\begin{figure}[h]
\includegraphics[width=0.6\textwidth]{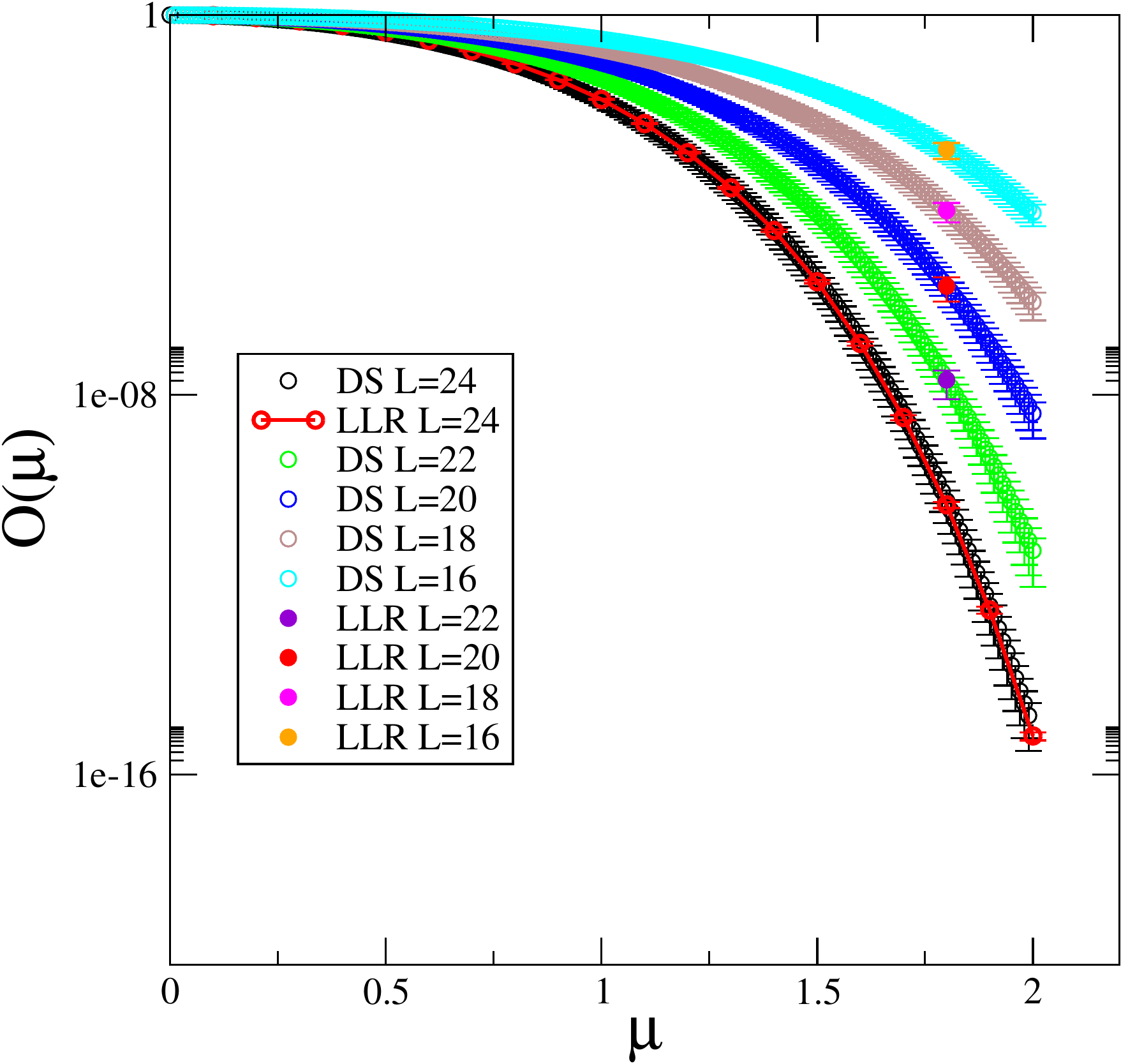}
\caption{\label{fig:4} The overlap $O(\mu)$ (\ref{eq:41}) from a
  direct Monte-Carlo simulation of the dual theory (DS) and from the
  LLR approach for several lattice sizes $L^3$. $\tau = 0.1$, $\kappa
  = 0.01$. }
\end{figure}
In order to quantify the influence of the imaginary part, we
introduce the partition function $Z_\mathrm{mod}$ that features the
$Z_3$ action (\ref{eq:10}) from which we have dropped the imaginary
part: 
\be 
S_\mathrm{mod}[z] = { \tau} \sum _{x,\nu} [z_x z^\ast _{x+\nu} +cc ] \; + \; 
\kappa \, \cosh (\mu) \, [ 2 N_1 \, + \, N_+ \, + \, N_- \, ] \; , 
\label{eq:40}
\en 
where $N_1$ is the number of $z=1$ elements on the lattice. 
The partition function of the modified theory does depend on the
chemical potential, but note that it can be simulated by standard
Monte-Carlo techniques since its Gibbs factor is real and positive by
construction. We then define the {\it overlap} between the full theory
and the modified theory by 
\be 
O(\mu ) = \frac{ Z(\mu) }{ Z_\mathrm{mod}(\mu) } \; . 
\label{eq:41}
\en 
We point out that being able to calculate the overlap $O(\mu)$
provides access to the observables of the full theory. For instance,
the density $\rho (\mu)$ acquires two parts: 
\be 
\rho (\mu) \; = \; \frac{d \, \ln Z(\mu) }{ d\mu } \; = \; 
\frac{d \, \ln O(\mu) }{ d\mu } \; + \; \frac{d \, \ln
  Z_\mathrm{mod}(\mu) }{ d\mu }  \; , 
\label{eq:42}
\en 
where the latter part is free of a sign problem and is calculable by
standard Monte-Carlo simulations. 

\medskip
An appealing feature of the $Z_3$ spin model is that its dual is a
real theory open for Monte-Carlo simulations. In fact, the theory can
be efficiently simulated by a worm type algorithm 
(see~\cite{Mercado:2011ua}) where the ``worms'' are conserved flux
lines of the dual theory (see~\cite{Langfeld:2013kno} for this
interpretation). It is still difficult to calculate the partition
function itself since theories with different chemical potentials
differ in the free energy density $f(\mu)$ leading to poor overlap: 
\be 
\frac{ Z(\mu + \Delta \mu) }{ Z(\mu) } \; = \; 
\exp \Bigl\{ - \, [ f(\mu + \Delta \mu) \, - \, f(\mu) ] \; V \Bigr\}
\; . 
\label{eq:43}
\en 
We used a variant of the ``snake
algorithm''~\cite{deForcrand:2000fi} to calculate ratios of the
partition function and to reconstruct the partition function from those: 
\be 
Z( k \, \Delta \mu ) \; = \; Z(0) \, \prod _{\ell=1}^k \frac{ Z(\ell \Delta
  \mu ) }{   Z((\ell-1) \Delta \mu ) } \; , 
\label{eq:44}
\en 
where $\Delta \mu $ must be chosen small enough (depending on the
number of degrees of freedom $V$) to ensure a good enough
signal-to-noise ratio. We here point out an advantage of the LLR
approach: $k$ simulations of the dual theories are necessary to
arrive at the target value $\mu _t \; = \; k \Delta \mu
$, while the LLR approach can aim directly at the chemical potential
$\mu _t$ of interest. 

\medskip 
We have simulated the $Z_3$ theory with non-zero chemical potentials
using the LLR method~\cite{Langfeld:2014nta}. We point out that the
method is {\it exact} implying that the numerical results need to
agree within error bars with the exact values. Note, however, that
sophisticated techniques for the error analysis (e.g.~bootstrap)  might
be needed and that standard Gaussian error analysis might fail at least
in certain regions of parameter space~\cite{Mercado:2014dva}.   
Our results from the direct simulation (DS) of the dual theory are 
shown in figure~\ref{fig:4} in comparison with our results from the
LLR approach. In the latter case, we have used
the method of the {\it polynomial fit} from subsection~\ref{sec:pf}
for the evaluation of the highly oscillating integral. We indeed
encounter a strong sign problem since the overlap is as small as
$10^{-16}$ for $\mu \approx 2$. So far, we have only explored a
limited region of the parameter space $(\tau,\kappa)$, but our results
serve as a proof of concept: for a range of volumes and for the
case of a strong sign problem, the simulation of the theory in its
original degrees of freedom is feasible using the LLR techniques. 

\section{Conclusions} 

Quantum field theory at finite density or, more general, statistical
systems with complex actions such as the imbalanced Fermi
gas~\cite{Goulko:2010rw} still await first principles results from
computer simulations. In the latter case, theoretical findings can be
scrutinised against experiments, and, given the level of abstraction that
went into model building, agreement would signal an 
understanding of the materials at hand (see
e.g.~\cite{Wingate:2012jh}). In the context of QCD at finite baryon
densities, a variety of mechanisms have been proposed over the last
couple of years that should describe the states of baryon matter in
the intermediate temperature and density range with or without a
strong magnetic field. Proposals feature ``quarkyonic
matter''~\cite{McLerran:2007qj}, suggested on the basis of the
't~Hooft limit, the ``chiral magnetic effect''~\cite{Fukushima:2010fe}
or the ``Fermi-Einstein condensation'' of
quarks~\cite{Langfeld:2011rh}, and this is not meant to be a complete
list. Lattice simulations would provide a genuine non-perturbative
approach with good systematic control of the errors, but are hampered
by the notorious sign problem: for a non-vanishing chemical potential,
the Gibbs factor is complex (or, at least, not positive definite) and
the action based importance sampling, which is at the heart of the
Monte-Carlo simulations, are impossible. 

\medskip 
Alongside the new theory proposals for the potential state of matter a
finite densities, the last decade has seen promising progress for the
simulation of theories with a sign problem. In fact, many of the
related ideas are rooted in the literature for decades, but techniques
have reached an unprecedented level of sophistication. A good example
are the Langevin simulation of complex actions systems, which date back
to the early works by Parisi~\cite{Parisi:1984cs} and Karsch and
Wyld~\cite{Karsch:1985cb} from the mid eighties, but underwent a
Renaissance when it was realised the Silver-Blaze problem can be
avoided for the case of a relativistic Bose gas~\cite{Aarts:2008wh}. 

\medskip 
Similarly, the LLR algorithm~\cite{Langfeld:2012ah} emerged from a
modification of Wang-Landau type algorithms~\cite{WangLandau2001} and
have progressed along the lines of the so-called density-of-states
methods (see e.g.~\cite{Bazavov:2012ex}
or~\cite{Azcoiti:1989rv,Anagnostopoulos:2001yb,Azcoiti:2002vk,Azcoiti:2011ei}). 
The  LLR algorithm copes with 
continuous degrees of freedom and is designed to numerically calculate
the probability distribution of an extensive quantity,
the action~\cite{Langfeld:2012ah} or e.g.~the Polyakov
line~\cite{Langfeld:2013xbf}, to an unprecedented precision and for
regions of the variable (e.g.~action) that would never be visited by an
action based importance sampling Monte-Carlo approach. Thus, the LLR
algorithm naturally solves overlap problems. Given the belief of a
correspondence between the overlap and the sign problem in finite
density quantum field theory, it was natural to explore its readiness
for complex action theories~\cite{Langfeld:2014nta}. Here, the LLR
algorithm provides a high quality probability distribution for the
imaginary part of the action, and the partition function emerges as
the Fourier transform of this distribution with the chemical potential
as its frequency (see (\ref{eq:3})). Details and advances of the LLR
methods have e.g.~been reported
in~\cite{Langfeld:2012ah,Pellegrini:2014gha,Mercado:2014dva,Langfeld:2014nta}. 
In this paper, we have focused on possible techniques to extract an
signal, which is exponentially small with the volume, from the highly
oscillating Fourier integral. As a proof of concept, we have studied
the $Z_3$ spin model~\cite{Langfeld:2014nta}. For this model,  we have
used (for a limited range of the parameter space) the Polynomial Fit
technique from subsection 3.1. Despite of a severe sign problem (as
quantified by a phase factor expectation value at the $10^{−16}$
level; see figure~\ref{fig:4}), we were able to obtain reliable
results by simulating the theory in its original formulation using the
LLR techniques.  An analysis of the full parameter
space of the $Z_3$ model, the LLR simulation of more involved theories
(e.g.~the $O(2)$-model) and the exploration of the techniques outlined
in section~\ref{sec:osc} to carry out the Fourier transform are
currently work in progress. 
 
\ack
We are indebted to Arieh Iserles for fruitful discussions on highly
oscillating integrals. This work is supported by STFC under the DiRAC
framework. We are grateful for the support from the HPCC Plymouth,
where the numerical computations have been carried out. KL and AR are
supported by the  Leverhulme Trust (grant RPG-2014-118) and STFC
(grant ST/L000350/1). BL is  supported by the Royal Society (grant
UF09003) and by STFC (grant ST/G000506/1).


\section*{References}
\begin{thebibliography}{9}

\bibitem{Langfeld:2012ah}
  K.~Langfeld, B.~Lucini and A.~Rago,
  Phys.\ Rev.\ Lett.\  {\bf 109} (2012) 111601
  [arXiv:1204.3243 [hep-lat]].


\bibitem{Laermann:2003cv}
  E.~Laermann and O.~Philipsen,
  Ann.\ Rev.\ Nucl.\ Part.\ Sci.\  {\bf 53} (2003) 163
  [hep-ph/0303042].

\bibitem{McLerran:2007qj}
  L.~McLerran and R.~D.~Pisarski,
  Nucl.\ Phys.\ A {\bf 796} (2007) 83
  [arXiv:0706.2191 [hep-ph]].

\bibitem{Fukushima:2010fe}
  K.~Fukushima, M.~Ruggieri and R.~Gatto,
  Phys.\ Rev.\ D {\bf 81} (2010) 114031
  [arXiv:1003.0047 [hep-ph]].

\bibitem{Langfeld:2011rh}
  K.~Langfeld and A.~Wipf,
  Annals Phys.\  {\bf 327} (2012) 994
  [arXiv:1109.0502 [hep-lat]].

\bibitem{Splittorff:2006fu}
  K.~Splittorff and J.~J.~M.~Verbaarschot,
  Phys.\ Rev.\ Lett.\  {\bf 98} (2007) 031601
  [hep-lat/0609076].

\bibitem{Fodor:2001au}
  Z.~Fodor and S.~D.~Katz,
  Phys.\ Lett.\ B {\bf 534} (2002) 87
  [hep-lat/0104001].

\bibitem{Fodor:2001pe}
  Z.~Fodor and S.~D.~Katz,
  JHEP {\bf 0203} (2002) 014
  [hep-lat/0106002].

\bibitem{Parisi:1984cs}
  G.~Parisi,
  Phys.\ Lett.\ B {\bf 131} (1983) 393.

\bibitem{Karsch:1985cb}
  F.~Karsch and H.~W.~Wyld,
  Phys.\ Rev.\ Lett.\  {\bf 55} (1985) 2242.

\bibitem{Aarts:2008wh}
  G.~Aarts,
  Phys.\ Rev.\ Lett.\  {\bf 102} (2009) 131601
  [arXiv:0810.2089 [hep-lat]].

\bibitem{Aarts:2009hn}
  G.~Aarts,
  JHEP {\bf 0905} (2009) 052
  [arXiv:0902.4686 [hep-lat]].

\bibitem{Aarts:2009uq}
  G.~Aarts, E.~Seiler and I.~O.~Stamatescu,
  Phys.\ Rev.\ D {\bf 81} (2010) 054508
  [arXiv:0912.3360 [hep-lat]].

\bibitem{Aarts:2014bwa}
  G.~Aarts, E.~Seiler, D.~Sexty and I.~O.~Stamatescu,
  Phys.\ Rev.\ D {\bf 90} (2014) 11,  114505
  [arXiv:1408.3770 [hep-lat]].

\bibitem{Rossi:1984cv}
  P.~Rossi and U.~Wolff,
  Nucl.\ Phys.\ B {\bf 248} (1984) 105.

\bibitem{deForcrand:2009dh}
  P.~de Forcrand and M.~Fromm,
  Phys.\ Rev.\ Lett.\  {\bf 104} (2010) 112005
  [arXiv:0907.1915 [hep-lat]].

\bibitem{Bloch:2013ara}
  J.~Bloch, F.~Bruckmann and T.~Wettig,
  JHEP {\bf 1310} (2013) 140
  [arXiv:1307.1416 [hep-lat]].

\bibitem{Mercado:2011ua}
  Y.~D.~Mercado, H.~G.~Evertz and C.~Gattringer,
  Phys.\ Rev.\ Lett.\  {\bf 106} (2011) 222001
  [arXiv:1102.3096 [hep-lat]].

\bibitem{Banerjee:2010kc}
  D.~Banerjee and S.~Chandrasekharan,
  Phys.\ Rev.\ D {\bf 81} (2010) 125007
  [arXiv:1001.3648 [hep-lat]].

\bibitem{Langfeld:2013kno}
  K.~Langfeld,
  Phys.\ Rev.\ D {\bf 87} (2013) 11,  114504
  [arXiv:1302.1908 [hep-lat]].

\bibitem{Prokof'ev:2009xw}
  N.~Prokof'ev and B.~Svistunov,
  Understanding Quantum Phase Transitions, Lincoln D. Carr,ed. Taylor \& Francis, Boca Raton, 2010
  [arXiv:0910.1393 [cond-mat.stat-mech]].

\bibitem{Brandt:2014rca} 
  B.~B.~Brandt and T.~Wettig,
  arXiv:1411.3350 [hep-lat].

\bibitem{Meurice:2014tca} 
  Y.~Meurice, Y.~Liu, J.~Unmuth-Yockey, L.~P.~Yang and H.~Zou,
  arXiv:1411.3392 [hep-lat].

\bibitem{Chandrasekharan:2009wc}
  S.~Chandrasekharan,
  Phys.\ Rev.\ D {\bf 82} (2010) 025007
  [arXiv:0910.5736 [hep-lat]].

\bibitem{Chandrasekharan:2011mn}
  S.~Chandrasekharan and A.~Li,
  Phys.\ Rev.\ Lett.\  {\bf 108} (2012) 140404
  [arXiv:1111.7204 [hep-lat]].

\bibitem{WangLandau2001}
 F.~Wang and D.~P.~Landau, 
 Phys.\ Rev.\ Lett.\ {\bf 86} (2001) 2050 
 [http://arxiv.org/abs/cond-mat/0011174] 

\bibitem{Bazavov:2012ex}
  A.~Bazavov, B.~A.~Berg, D.~Du and Y.~Meurice,
  Phys.\ Rev.\ D {\bf 85} (2012) 056010
  [arXiv:1202.2109 [hep-lat]].

\bibitem{Pellegrini:2014gha}
  R.~Pellegrini, K.~Langfeld, B.~Lucini and A.~Rago,
  arXiv:1411.0655 [hep-lat].

\bibitem{Langfeld:2013xbf}
  K.~Langfeld and J.~M.~Pawlowski,
  Phys.\ Rev.\ D {\bf 88} (2013) 7,  071502
  [arXiv:1307.0455 [hep-lat]].

\bibitem{Langfeld:2014nta}
  K.~Langfeld and B.~Lucini,
  Phys.\ Rev.\ D {\bf 90} (2014) 9,  094502
  [arXiv:1404.7187 [hep-lat]].

\bibitem{Aarts:2015kea}
  G.~Aarts,
  PoS CPOD {\bf 2014} (2014) 012
  [arXiv:1502.01850 [hep-lat]].

\bibitem{deForcrand:2000fi}
  P.~de Forcrand, M.~D'Elia and M.~Pepe,
  Phys.\ Rev.\ Lett.\  {\bf 86} (2001) 1438
  [hep-lat/0007034].

\bibitem{Mercado:2014dva}
  Y.~D.~Mercado, P.~Törek and C.~Gattringer,
  arXiv:1410.1645 [hep-lat].

\bibitem{Goulko:2010rw}
  O.~Goulko and M.~Wingate,
  PoS LATTICE {\bf 2010} (2010) 187
  [arXiv:1011.0312 [cond-mat.quant-gas]].

\bibitem{Wingate:2012jh}
  M.~Wingate and O.~Goulko,
  PoS LATTICE {\bf 2012} (2012) 057.

\bibitem{Gocksch1988}
 A.~Gocksch, 
Phys.\ Rev.\ Lett.\ 61 (1988) 2054. 

\bibitem{Azcoiti:1989rv}
  V.~Azcoiti, A.~Cruz, A.~Tarancon and G.~Di Carlo,
  Nucl.\ Phys.\ Proc.\ Suppl.\  {\bf 17} (1990) 727.

\bibitem{Anagnostopoulos:2001yb}
  K.~N.~Anagnostopoulos and J.~Nishimura,
  Phys.\ Rev.\ D {\bf 66} (2002) 106008
  [hep-th/0108041].

\bibitem{Azcoiti:2002vk}
  V.~Azcoiti, G.~Di Carlo, A.~Galante and V.~Laliena,
  Phys.\ Rev.\ Lett.\  {\bf 89} (2002) 141601
  [hep-lat/0203017].

\bibitem{Azcoiti:2011ei}
  V.~Azcoiti, E.~Follana and A.~Vaquero,
  Nucl.\ Phys.\ B {\bf 851} (2011) 420
  [arXiv:1105.1020 [hep-lat]].



\end{thebibliography}
\end{document}